\documentclass{PoS}
\usepackage{cite}
\usepackage{amsmath}
\usepackage{eufrak}
\usepackage{amssymb}
\usepackage{eufrak}
\newcommand{\gsim}{\raisebox{-0.07cm  }
{$\, \stackrel{>}{{\scriptstyle\sim}}\, $}}

\newcommand{\ep}{\varepsilon}
\newcommand{\N}{\nonumber}

\newcommand{\MS}{\overline{{\sf MS}}}

\graphicspath{{Pics/}}
\DeclareGraphicsExtensions{.eps,.ps}
\title{\small{DESY 10-065, DO-TH 12/05, SFB/CPP-12-09, LPN 12-036   \hfill 
}
\\ \LARGE
New Heavy Flavor Contributions to the DIS Structure Function $F_2(x,Q^2)$ at 
\boldmath{$O(\alpha_s^3)$}}

\ShortTitle{New Heavy Flavor Contributions ...}

\author{J.~Ablinger$^{a,b}$,  \speaker{J.~Bl\"umlein}$^a$, A.~Hasselhuhn$^a$, S.~Klein$^c$, C.~Schneider$^b$, 
F.~Wi\ss{}brock$^a$\\
\llap{$^a$}DESY, Zeuthen, Platanenalle 6, D-15735 Zeuthen, Germany.\\
\llap{$^b$}RISC, Johannes Kepler Universit\"at Linz, Altenberger Str. 69, A-4040 Linz, Austria
\\
\llap{$^c$}Inst. Theor. Teilchenphysik und Kosmologie, RWTH Aachen University, D-52056 Aachen, Germany.}

\abstract{We report on recent results obtained for the massive Wilson coefficients which contribute 
to the structure function $F_2(x,Q^2)$ at $O(\alpha_s^3)$ in the region $Q^2/m^2 \gsim 10$. In the 
calculation new species of harmonic sums and harmonic polylogarithms generated by cyclotomic 
polynomials arise in intermediary results which are briefly discussed.} 

\FullConference{
10th International Symposium on Radiative Corrections (Applications of Quantum Field Theory to Phenomenology), RADCOR 
2011
Mamallapuram, India
September 26-30, 2011}

\begin{document}

\section{Introduction}

\noindent
The charm quark contributions to the nucleon structure functions are large in the small $x$ region.
To obtain a correct theoretical description of the structure function $F_2(x,Q^2)$, which is measured 
at an accuracy of $1\%$ in this region, 3-loop (NNLO) corrections to the heavy flavor Wilson coefficients
are needed. Present extractions of the strong coupling constant from deep inelastic data \cite{ALPH} and 
precision measurements of parton densities require this accuracy, with important consequences for 
precision measurements at Tevatron and LHC \cite{Alekhin:2010dd}. In previous calculations the Wilson 
coefficients were obtained to $O(\alpha_s^2)$ \cite{NLO}~\footnote{Precise numerical implementations in 
Mellin space were given in \cite{Alekhin:2003ev}.}. If one considers the region $Q^2/m^2 \gsim 10$ the 
massive Wilson coefficients can be obtained through convolutions \cite{Buza:1995ie,Bierenbaum:2009mv} of 
massive operator matrix elements (OMEs) \cite{Bierenbaum:2007zz,Bierenbaum:2007qe,Bierenbaum:2007dm,
Bierenbaum:2008yu,Bierenbaum:2009zt,Ablinger:2010ty,Ablinger:2011pb} and the massless Wilson coefficients 
\cite{Vermaseren:2005qc}. Other massive OMEs play a role in higher order QED correction, \cite{Blumlein:2011mi}.
Based on the 2-loop massive operator matrix elements, which are known for general values of $N$  
up to the linear term in the dimensional parameter $\varepsilon$ \cite{Bierenbaum:2007qe,Bierenbaum:2008yu,
Bierenbaum:2009zt} and the massless 3-loop Wilson coefficients \cite{Vermaseren:2005qc} one may compute all 
logarithmic contributions at 3--loop order \cite{Bierenbaum:2010jp}. In the kinematic region of HERA, however, 
the logarithmic terms are not dominant over the yet unknown constant terms for general values of $N$, as 
demonstrated for the moments calculated in \cite{Bierenbaum:2009mv}. The $O(\alpha_s^3)$ corrections to the 
structure function $F_L(x,Q^2)$ in the asymptotic region were computed in 
\cite{Blumlein:2006mh,Bierenbaum:2009mv}.
Heavy flavor corrections for charged current reactions in Mellin space were recently given in \cite{CC},
correcting some errors in the literature.

The heavy flavor contributions to the structure function $F_2(x,Q^2)$ at $O(\alpha_s^3)$ in case of one 
massive quark are described by five massive Wilson coefficients $L_q^{\rm NS}, L_q^{\rm PS},
L_g^{\rm S}, H_q^{\rm PS}, H_q^{\rm S}$ \cite{Bierenbaum:2009mv} in the asymptotic region. Two of these 
Wilson coefficients,  $L_q^{\rm PS}$ and $L_g^{\rm S}$ haven been computed completely for general values of 
$N$ in \cite{Ablinger:2010ty}, cf. also \cite{Bierenbaum:2010jp}. In \cite{Ablinger:2010ty} the 
contributions of the color factors $O(T_F^2 N_f C_{A,F})$ to the Wilson coefficients $L_q^{\rm NS}, 
H_q^{\rm PS}, H_q^{\rm S}$ were also calculated. The corresponding Feynman diagrams consist of graphs with 
a massless and a massive internal fermion line. After applying algebraic relations 
\cite{ALGEBRA} these contributions to the massive Wilson coefficients can be represented in terms of 
the known {\sf w = 4} set of harmonic sums \cite{BRK}. The calculations of the massive OMEs at general 
value of $N$ were performed using {\tt tform} \cite{TFORM} and C. Schneider's summation package {\tt Sigma} 
\cite{SIGMA}, which has received continuous updates during the last years.

In this note we report on further progress of the 3-loop calculation of massive OMEs on contributions 
due to charm and bottom lines in single diagrams, massive gluonic OMEs, massive 3--loop ladder graphs,
and cyclotomic harmonic sums and polylogarithms, which emerge in intermediate steps of present 
calculations.
\section{Two massive quarks of unequal mass}

\noindent
Beginning at $O(\alpha_s^3)$, graphs with internal massive fermion lines carrying unequal finite masses
contribute. Due to the mass ratio $x =m_c^2/m_b^2 \simeq 1/10$ for the case of charm and bottom   
quarks, one may expand the corresponding diagrams using this parameter. The results for the
moments $N=2$ and $N=4$ of the gluonic operator matrix element $A_{Qg}$ were given in  
\cite{Ablinger:2011pb} before, extending the code {\tt qexp} \cite{QEXP} to higher moments applying 
projectors similar to those used in \cite{Bierenbaum:2009mv}. With rising values of $N$ the calculations 
become more and more time consuming even using {\tt tform} \cite{TFORM}. Here we present the moment $N=6$
which requested several months of run time~:  

{{\footnotesize
\begin{eqnarray}
a_{Qg}^{(3)}&=&
        T_F^2 C_A \Bigg\{
          \frac{69882273800453}{367569090000}
          - \frac{395296}{19845} \zeta_3
          + \frac{1316809}{39690} \zeta_2
          + \frac{832369820129}{14586075000} x
          + \frac{1511074426112}{624023544375} x^2
\N \\&&
          - \frac{84840004938801319}{690973782403905000} x^3
\N \\&&
       + \ln\Bigl( \frac{m_2^2}{\mu^2} \Bigr)    \Bigl[
         \frac{11771644229}{194481000}
          + \frac{78496}{2205} \zeta_2
          - \frac{1406143531}{69457500} x
          - \frac{105157957}{180093375} x^2
          + \frac{2287164970759}{7669816654500} x^3
          \Bigr]
\N\\&&
       + \ln^2\Bigl(\frac{m_2^2}{\mu^2}\Bigr)    \Bigl[
          \frac{2668087}{79380}
          + \frac{112669}{661500} x
          - \frac{49373}{51975} x^2
          - \frac{31340489}{34054020} x^3
          \Bigr]
             + \ln^3\Bigl(\frac{m_2^2}{\mu^2}\Bigr) \frac{324148}{19845}
\N\\&&
+ \ln^2\Bigl(\frac{m_2^2}{\mu^2}\Bigr)
       \ln\Bigl(\frac{m_1^2}{\mu^2}\Bigr)   \frac{156992}{6615}
\N\\&&
       + \ln\Bigl(\frac{m_2^2}{\mu^2}\Bigr)
       \ln\Bigl(\frac{m_1^2}{\mu^2}\Bigr)    \Bigl[
       \frac{128234}{3969}
          -  \frac{112669}{330750} x
          +  \frac{98746}{51975} x^2
          +  \frac{31340489}{17027010} x^3
          \Bigr]
       + \ln\Bigl(\frac{m_2^2}{\mu^2}\Bigr)
       \ln^2\Bigl(\frac{m_1^2}{\mu^2}\Bigr) \frac{68332}{6615}
\N\\&&
       + \ln\Bigl(\frac{m_1^2}{\mu^2}\Bigr) \Bigl[
          \frac{83755534727}{583443000}
          + \frac{78496}{2205} \zeta_2
          + \frac{1406143531}{69457500} x
          + \frac{105157957}{180093375} x^2
          - \frac{2287164970759}{7669816654500} x^3
          \Bigr]
\N\\&&
       + \ln^2\Bigl(\frac{m_1^2}{\mu^2}\Bigr)  \Bigl[
           \frac{2668087}{79380}
          + \frac{112669}{661500} x
          - \frac{49373}{51975} x^2
          - \frac{31340489}{34054020} x^3
          \Bigr]
       + \ln^3\Bigl(\frac{m_1^2}{\mu^2}\Bigr) \frac{412808}{19845}
\Biggr\}
\N\\&&
+T_F^2 C_F \Bigg\{
          - \frac{3161811182177}{71471767500}
          + \frac{447392}{19845} \zeta_3
          + \frac{9568018}{4862025} \zeta_2
          - \frac{64855635472}{2552563125} x
          + \frac{1048702178522}{97070329125} x^2
\N \\&&
          + \frac{1980566069882672}{2467763508585375} x^3
\N\\&&
       + \ln\Bigl(\frac{m_2^2}{\mu^2}\Bigr)    \Bigl[
          \frac{1786067629}{204205050}
          - \frac{111848}{15435} \zeta_2
          - \frac{128543024}{24310125} x
          - \frac{22957168}{3361743} x^2
          - \frac{2511536080}{2191376187} x^3
          \Bigr]
\N\\&&
              + \ln^2\Bigl(\frac{m_2^2}{\mu^2}\Bigr)    \Bigl[
          \frac{3232799}{4862025}
          + \frac{752432}{231525} x
          + \frac{177944}{40425} x^2
          + \frac{127858928}{42567525} x^3
          \Bigr]
       - \ln^3\Bigl(\frac{m_2^2}{\mu^2}\Bigr)    \frac{111848}{19845}
\N\\&&
       - \ln^2\Bigl(\frac{m_2^2}{\mu^2}\Bigr)
       \ln\Bigl(\frac{m_1^2}{\mu^2}\Bigr)      \frac{223696}{46305}
\N\\&&
       + \ln\Bigl(\frac{m_2^2}{\mu^2}\Bigr)
       \ln\Bigl(\frac{m_1^2}{\mu^2}\Bigr)    \Bigl[
            \frac{22238456}{4862025}
          - \frac{1504864}{231525} x
          - \frac{355888}{40425} x^2
          - \frac{255717856}{42567525} x^3
          \Bigr]
       + \ln\Bigl(\frac{m_2^2}{\mu^2}\Bigr)
       \ln^2\Bigl(\frac{m_1^2}{\mu^2}\Bigr) \frac{223696}{46305}
\N\\&&
       + \ln\Bigl(\frac{m_1^2}{\mu^2}\Bigr)    \Bigl[
                 - \frac{24797875607}{1021025250}
          - \frac{111848}{15435} \zeta_2
          + \frac{128543024}{24310125} x
          + \frac{22957168}{3361743} x^2
          + \frac{2511536080}{2191376187} x^3
          \Bigr]
\N\\&&
       + \ln^2\Bigl(\frac{m_1^2}{\mu^2}\Bigr)    \Bigl[
          \frac{3232799}{4862025}
          + \frac{752432}{231525} x
          + \frac{177944}{40425} x^2
          + \frac{127858928}{42567525} x^3
          \Bigr]
       - \ln^3\Bigl(\frac{m_1^2}{\mu^2}\Bigr) \frac{1230328}{138915}
\Biggr\}+O\left(x^4 \ln^3(x)\right).
\N\\
\label{eq:MOM6}
\end{eqnarray}
}}

\noindent
These contributions to the massive Wilson coefficients can be uniquely calculated in the fixed flavor 
number scheme with three massless quarks in the initial state. However they cannot be attributed either
to the charm- or bottom distribution in a variable flavor scheme.~\footnote{It was shown in \cite{BVN} that 
the matching scale is rather process dependent and may differ significantly from the heavy quark mass 
$m_Q$, as comparisons to complete calculations show. This aspect, however, is often ignored.}
Despite all these contributions are universal, the so-called variable flavor scheme {\it ends at the 2-loop
level.} 
In general we remark that the representation of the heavy flavor Wilson coefficients to 
$O(\alpha_s^3)$ are given in the fixed flavor number scheme. As has been shown e.g. in Refs.~\cite{DO,ABKM} 
this choice is sufficient for the kinematic range at HERA.
\section{Contributions due to 3-Loop Ladder Graphs} 

\noindent 
A next topology class consists of ladder graphs containing up to six massive lines and local operator 
insertions. The calculation of these graphs is underway \cite{ABHKSW}. Here we present, as an example, the 
result for the scalar Diagram shown in the following Figure~:
\begin{figure}[h]
\begin{center}
\includegraphics[scale=.7]{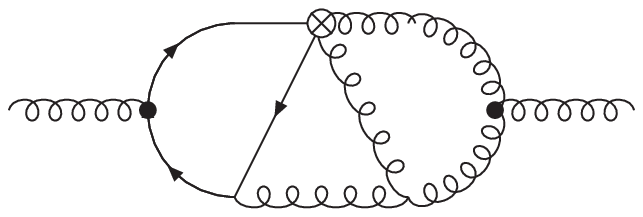}
\end{center}
\end{figure}

\vspace*{-12mm}
\begin{eqnarray}
\hat{I}_{10a} &=&  
   \frac{1}{(N+3) (N+4)}
   \Biggl\{
     \Biggl[
   -\frac{4 \left(N^3+3 N^2-N-5\right)}{(N+1) (N+2) (N+3)} S_{1}
   +2 S_{1}^2
   +\frac{4 (-1)^N}{N+3} S_{1}
   +4 S_{-2}
\N\\&&
   +2 (2 N+5) S_{2}
   +\frac{4 (-1)^N \left(2 N^3+7 N^2+4 N-3\right)}{(N+1)^2 (N+2)^2 (N+3)}
   +\frac{4 \left(6 N^3+34 N^2+63 N+39\right)}{(N+1)^2 (N+2)^2 (N+3)}
      \Biggr]
      \frac{1}{\ep^2}
\N\\&&
     +\Biggl[
    \frac{\left(-4 N^4-25 N^3-30 N^2+49 N+76\right)}{(N+1) (N+2) (N+3) (N+4)} S_{1}^2
   -\frac{4 \left(2 N^4+14 N^3+27 N^2+5 N-16\right)}{(N+1) (N+2) (N+3) (N+4)} S_{-2}
\N\\
&&
   +\frac{\left(10 N^4+73 N^3+158 N^2+73 N-52\right)}{(N+1) (N+2) (N+3) (N+4)} S_{2}
\N\\&&
   +\frac{2 (-1)^N \left(12 N^5+127 N^4+538 N^3+1177 N^2+1354 N+648\right)}{(N+1)^2 (N+2)^2 (N+3)^2 (N+4)} S_{1}
\N\\&&
   -\frac{2 \left(8 N^6+51 N^5-72 N^4-1330 N^3-4062 N^2-5151 N-2436\right)}{(N+1)^2 (N+2)^2 (N+3)^2 (N+4)} S_{1}
\N\\&&
   +S_{1}^3
   +\frac{(-1)^N}{N+3}\left(
      S_{1}^2
     -S_{2}
   \right)
   +4 S_{-2} S_{1}
   -5 S_{2} S_{1}
   +2 (4 N+15) S_{-3}
   +2 (N-1) S_{3}
\N\\&&
   -12 S_{-2,1}
   +8 (N+4) S_{2,1}
\N\\&&
   +\frac{2 (-1)^N \left(11 N^6+60 N^5-160 N^4-1837 N^3-5005 N^2-5801 N-2508\right)}{(N+1)^3 (N+2)^3 (N+3)^2 (N+4)}
\N\\&&
   +\frac{2 \left(70 N^6+893 N^5+4640 N^4+12626 N^3+19074 N^2+15269 N+5100\right)}{(N+1)^3 (N+2)^3 (N+3)^2 (N+4)}
     \Biggr]
     \frac{1}{\ep}
\N\\
& &
   +\frac{7}{24} S_{1}^4
   +\frac{\left(-10 N^4-61 N^3-68 N^2+129 N+188\right)}{6 (N+1) (N+2) (N+3) (N+4)} S_{1}^3
\N\\&&
   +\frac{(-1)^N \left(12 N^5+127 N^4+538 N^3+1177 N^2+1354 N+648\right)}{2 (N+1)^2 (N+2)^2 (N+3)^2 (N+4)} S_{1}^2
\N\\&&
   +\frac{P_{22}}{2 (N+1)^2 (N+2)^2 (N+3)^2 (N+4)^2} S_{1}^2
   +\frac{3}{4} \zeta_2 S_{1}^2
   -4 S_{-2} S_{1}^2
   -\frac{13}{4} S_{2} S_{1}^2
\N
\end{eqnarray}
\begin{eqnarray}
&&
   +\frac{(-1)^N P_{23}}{(N+1)^3 (N+2)^3 (N+3)^3 (N+4)^2} S_{1}
   +\frac{P_{24}}{(N+1)^3 (N+2)^3 (N+3)^3 (N+4)^2} S_{1}
\N\\&&
   -\frac{3 \left(N^3+3 N^2-N-5\right)}{2 (N+1) (N+2) (N+3)} \zeta_2 S_{1}
   -2 S_{-3} S_{1}
\N\\&&
   -\frac{4 \left(4 N^4+41 N^3+155 N^2+254 N+148\right)}{(N+1) (N+2) (N+3) (N+4)} S_{-2} S_{1}
\N\\ &&
   +\frac{(-1)^N}{N+3}\left(
   -4 S_{-2} S_{1}
   +\frac{9}{2} S_{2} S_{1}
   +\frac{3}{2} \zeta_2 S_{1}
   +\frac{1}{6} S_{1}^3
   -2 S_{-3}
   +\frac{10}{3} S_{3}
   +2 S_{2,1}
   +12 S_{-2,1}
   \right)
\N\\&&
   +\frac{\left(-14 N^4-201 N^3-936 N^2-1715 N-1044\right)}{2 (N+1) (N+2) (N+3) (N+4)} S_{2} S_{1}
   -\frac{119}{3} S_{3} S_{1}
\N\\&&
   -12 S_{-2,1} S_{1}
   +22 S_{2,1} S_{1}
   -2 S_{-2}^2
   +\frac{1}{8} (32 N+119) S_{2}^2
\N\\&&
   +\frac{(-1)^N P_{25}}{(N+1)^4 (N+2)^4 (N+3)^3 (N+4)^2}
   +\frac{P_{26}}{(N+1)^4 (N+2)^4 (N+3)^3 (N+4)^2}
\N
\\
&&
   +\frac{3 (-1)^N \left(2 N^3+7 N^2+4 N-3\right)}{2 (N+1)^2 (N+2)^2 (N+3)} \zeta_2
   +\frac{3 \left(6 N^3+34 N^2+63 N+39\right)}{2 (N+1)^2 (N+2)^2 (N+3)} \zeta_2
\N\\
&&
   +(8 N+39) S_{-4}
   +\frac{2 \left(8 N^5+108 N^4+558 N^3+1365 N^2+1553 N+640\right)}{(N+1) (N+2) (N+3) (N+4)} S_{-3}
\N\\&&
   -\frac{4 (-1)^N \left(2 N^3+7 N^2+4 N-3\right)}{(N+1)^2 (N+2)^2 (N+3)} S_{-2}
   -\frac{4 P_{27}}{(N+1)^2 (N+2)^2 (N+3)^2 (N+4)^2} S_{-2}
\N\\&&
   +\frac{3}{2} \zeta_2 S_{-2}
   +\frac{(-1)^N \left(8 N^5+79 N^4+186 N^3-279 N^2-1426 N-1224\right)}{2 (N+1)^2 (N+2)^2 (N+3)^2 (N+4)} S_{2}
\N\\&&
   +\frac{P_{28}}{2 (N+1)^2 (N+2)^2 (N+3)^2 (N+4)^2} S_{2}
   +\frac{3}{4} (2 N+5) \zeta_2 S_{2}
   +8 S_{-2} S_{2}
\N\\
&&
   +\frac{\left(-18 N^5-229 N^4-1498 N^3-5558 N^2-10017 N-6460\right)}{3 (N+1) (N+2) (N+3) (N+4)} S_{3}
\N\\&&
   +\frac{1}{4} (20 N-29) S_{4}
   -14 S_{-3,1}
   +\frac{4 \left(4 N^4+22 N^3+11 N^2-85 N-96\right)}{(N+1) (N+2) (N+3) (N+4)} S_{-2,1}
\N\\&&
   -14 S_{-2,2}
   +\frac{2 \left(11 N^4+107 N^3+397 N^2+640 N+361\right)}{(N+1) (N+2) (N+3)} S_{2,1}
   +2 (N+36) S_{3,1}
\N\\&&
   +28 S_{-2,1,1}
   +2 (2 N-7) S_{2,1,1}\Biggr\}
  + O(\ep)~,
\\
P_{22} &=& -6 N^8-164 N^7-1613 N^6-7762 N^5-19526 N^4-22888 N^3-2137 N^2
\N\\&&
+19968 N+13264~,\\
P_{23} &=& 119 N^8+2250 N^7+18755 N^6+90365 N^5+275464 N^4+542281 N^3+668958 N^2
\N\\&&
+469072 N+142112~,\\
P_{24} &=& 16 N^{11}+448 N^{10}+5568 N^9+41171 N^8+204092 N^7+720291 N^6+1858328 N^5
\N\\&&
+3504939 N^4+4712624 N^3+4272331 N^2+2335952 N+581072~,
\\
P_{25} &=& 78 N^9+937 N^8+2466 N^7-17638 N^6-155141 N^5-538674 N^4-1047495 N^3
\N\\&&
-1197445 N^2-757472 N-206256~,
\end{eqnarray}
\begin{eqnarray}
P_{26} &=& 568 N^9+11297 N^8+98332 N^7+492027 N^6+1561688 N^5+3266831 N^4
\N\\&&
+4516420 N^3+3994885 N^2+2061840 N+475824~,\\
P_{27} &=& 4 N^8+96 N^7+942 N^6+4995 N^5+15753 N^4+30351 N^3+34903 N^2
\N\\&&
+21844 N+5648~,\\
P_{28} &=& -32 N^9-730 N^8-7180 N^7-40057 N^6-139918 N^5-317434 N^4-466820 N^3
\N\\&&
-426421 N^2-216416 N-45040~.
\end{eqnarray}
\section{Contributions to Gluonic Matrix Elements} 

\noindent 
First results for general values of $N$ have been obtained for the gluonic OMEs of $O(N_f T_F^2 C_{A,F}$. 
They are of importance to define the variable flavor scheme for the single heavy mass contributions, 
cf.~\cite{Buza:1996wv,Bierenbaum:2009mv,Bierenbaum:2009zt}. As an example we show the renormalized result 
for $A_{gq,Q}^{(3)}(N)$, cf.~\cite{BHS}:
\parbox[t]{\textwidth}{
\begin{eqnarray}
  A_{gq,Q,C_F T_F^2 n_f}^{(3), \MS}&=&
   C_F n_f T_F^2 \frac{1+(-1)^N}{2} \Biggl\{
  \frac{88 \left(N^2+N+2\right)}{27 (N-1) N (N+1)} 
  \log^3\left(\frac{m^2}{\mu^2}\right)
\N\\ &&
  +\left(
    \frac{8 \left(N^2+N+2\right)}{9 (N-1) N (N+1)} S_{1}
   -\frac{8 \left(8 N^3+13 N^2+27 N+16\right)}{27 (N-1) N (N+1)^2}
  \right) \log^2\left(\frac{m^2}{\mu^2}\right)
\N\\ &&
  +\Biggl[
    \frac{16 \left(N^2+N+2\right) }{9 (N-1) N (N+1)}
     \left( S_{1}^2 + S_{2} - \frac{1}{2}\zeta_2 \right)
   -\frac{32 \left(8 N^3+13 N^2+27 N+16\right) }{27 (N-1) N (N+1)^2} S_{1}
\N\\ &&
   +\frac{16 \left(179 N^4+489 N^3+913 N^2+925 N+358\right)}{81 (N-1) N (N+1)^3}
  \Biggr] \log\left(\frac{m^2}{\mu^2}\right)
\N\\ &&
  +\Biggl[
    \frac{16 (N^2+N+2)}{9 (N-1) N (N+1)}
     \left( S_{1}^3 + 3S_{2} S_{1} + 2S_{3} + 5S_{1} \zeta_2 + \frac{11}{2}\zeta_3\right)
\N\\ &&
   -\frac{16 (8 N^3+13 N^2+27 N+16)}{9 (N-1) N (N+1)^2} 
    \left(S_{1}^2 + S_{2} + \frac{5}{3}\zeta_2\right)
\N\\ &&
   +\frac{32 (160 N^4+408 N^3+827 N^2+845 N+320)}{81 (N-1) N (N+1)^3} S_{1}
\N\\ &&
   -\frac{32 (1189 N^5+4276 N^4+9248 N^3+12289 N^2+8668 N+2378)}{243 (N-1) N (N+1)^4}
  \Biggr] 
  \Biggr\}.
\end{eqnarray}
}
\section{Cyclotomic Harmonic Sums and Harmonic Polylogarithms}

\noindent
In various calculations of the massive OMEs harmonic sums emerged, which are associated to cyclotomic
harmonic polylogarithms. It appeared therefore as useful to generalize the usual harmonic polylogarithms
\cite{Remiddi:1999ew} based on the alphabet $\{1/x, 1/(1-x), 1/(1+x)\}$ to that of the alphabet
\begin{eqnarray}
\mathfrak{A} =
\left\{\frac{1}{x}\right\}\cup\left\{\left. \frac{x^l}{\Phi_k(x)}\right|k\in\mathbb
N_+,0\leq
l<\varphi(k)\right\},
\end{eqnarray}
with $\Phi_k$ the $k$th cyclotomic polynomial and $\varphi$ Euler's totient function, cf. 
\cite{Ablinger:2011te}. 
The cyclotomic harmonic polynomials are obtained as iterated integrals of the letters in $\mathfrak{A}$. 
The corresponding harmonic sums\footnote{For the definition of nested harmonic sums see \cite{HSUM}.} 
iterate the letters of the type
\begin{eqnarray}
\frac{(-1)^k}{(nk+m)^l},~~~n \geq 1, m \leq n, l \geq 1~,
\end{eqnarray}
which can also be related to usual harmonic sums at numerator weights being $n$th roots of unity.
The properties of these harmonic polylogarithms, harmonic sums and associated generalizations of the 
multiple zeta values \cite{MZV}, as well as the relations of these quantities, are given in 
\cite{Ablinger:2011te}. They are implemented in the package {\tt HarmonicSums} \cite{ADISS}. 

\vspace*{2mm}
\noindent
{\bf Acknowledgement.}
This work has been supported in part by Studienstiftung des Deutschen Volkes, 
DFG Sonderforschungsbereich Transregio 9, Computergest\"utzte Theoretische Teilchenphysik, Austrian Science 
Fund (FWF) grant P203477-N18, and EU Network {\sf LHCPHENOnet} PITN-GA-2010-264564. 



\begin{thebibliography}{99}
%
\bibitem{ALPH}
  J.~Bl\"umlein,
  Mod.\ Phys.\ Lett.\  {\bf A25 } (2010)  2621
  [arXiv:1007.5202 [hep-ph]];\\
  S.~Alekhin, J.~Bl\"umlein, H.~B\"ottcher, S.~-O.~Moch,
  [arXiv:1104.0469 [hep-ph]].
%
\bibitem{Alekhin:2010dd}
  S.~Alekhin, J.~Bl\"umlein, P.~Jimenez-Delgado, S.~Moch, E.~Reya,
  Phys. Lett. B {\bf B697 } (2011)  127
  [arXiv:1011.6259 [hep-ph]].
%
\bibitem{NLO}
  E.~Laenen, S.~Riemersma, J.~Smith, W.L. van Neerven,
  Nucl.\ Phys.\  {\bf B392 } (1993)  162;
229.\\
  S.~Riemersma, J.~Smith, W.~L.~van Neerven,
  Phys.\ Lett.\  {\bf B347 } (1995)  143,
  [hep-ph/9411431].
%
\bibitem{Alekhin:2003ev}
  S.~I.~Alekhin and J.~Bl\"umlein,
  Phys.\ Lett.\  B {\bf 594} (2004) 299,
  [arXiv:hep-ph/0404034].
%
\bibitem{Buza:1995ie}
  M.~Buza, Y.~Matiounine, J.~Smith, R.~Migneron and W.~L.~van Neerven,
  Nucl.\ Phys.\  B {\bf 472} (1996) 611,
  [hep-ph/9601302];\\
%
\bibitem{Bierenbaum:2009mv}
  I.~Bierenbaum, J.~Bl{\"u}mlein and S.~Klein,
  Nucl.\ Phys.\  B {\bf 820} (2009) 417,
   [hep-ph/0904.3563].
%
\bibitem{Bierenbaum:2007zz}
  I.~Bierenbaum, J.~Bl\"umlein and S.~Klein,
  PoS ACAT {\bf } (2007) 070.
%
\bibitem{Bierenbaum:2007qe}
  I.~Bierenbaum, J.~Bl\"umlein and S.~Klein,
  Nucl.\ Phys.\  B {\bf 780} (2007) 40,
  [hep-ph/0703285];\\
%
\bibitem{Bierenbaum:2007dm}
I.~Bierenbaum, J.~Bl\"umlein and S.~Klein,
  Phys.\ Lett.\  B {\bf 648} (2007) 195,
   [hep-ph/0702265];
%
\bibitem{Bierenbaum:2008yu}
  I.~Bierenbaum, J.~Bl\"umlein, S.~Klein and C.~Schneider,
  Nucl.\ Phys.\  B {\bf 803} (2008) 1,
  [hep-ph/0803.0273].
%
\bibitem{Bierenbaum:2009zt}
  I.~Bierenbaum, J.~Bl{\"u}mlein and S.~Klein,
  Phys.\ Lett.\  B {\bf 672} (2009) 401,
  [hep-ph/0901.0669].
%
\bibitem{Ablinger:2010ty}
  J.~Ablinger, J.~Bl\"umlein, S.~Klein, C.~Schneider and F.~Wissbrock,
  Nucl.\ Phys.\  B {\bf 844} (2011) 26
  [arXiv:1008.3347 [hep-ph]].
%
\bibitem{Ablinger:2011pb}
  J.~Ablinger, J.~Bl\"umlein, S.~Klein, C.~Schneider and F.~Wissbrock,
  arXiv:1106.5937 [hep-ph].
%
\bibitem{Vermaseren:2005qc}
  J.~A.~M.~Vermaseren, A.~Vogt and S.~Moch,
  Nucl.\ Phys.\  B {\bf 724} (2005) 3,
  [hep-ph/0504242].
%
\bibitem{Blumlein:2011mi}
  J.~Bl\"umlein, A.~De Freitas and W.~van Neerven,
  Nucl.\ Phys.\ B {\bf 855} (2012) 508
  [arXiv:1107.4638 [hep-ph]].
%
\bibitem{Bierenbaum:2010jp}
  I.~Bierenbaum, J.~Bl\"umlein and S.~Klein,
  arXiv:1008.0792 [hep-ph]. 
%
\bibitem{Blumlein:2006mh}
  J.~Bl\"umlein, A.~De Freitas, W.~L.~van Neerven and S.~Klein,
  Nucl.\ Phys.\  B {\bf 755} (2006) 272,
  [hep-ph/0608024].
%
\bibitem{CC}
  M.~Gl\"uck, S.~Kretzer, E.~Reya,
  Phys.\ Lett.\  {\bf B380 } (1996)  171
  [hep-ph/9603304];\\
  J.~Bl\"umlein, A.~Hasselhuhn, P.~Kovacikova and S.~Moch,
  Phys.\ Lett.\  B {\bf 700} (2011) 294
  [arXiv:1104.3449 [hep-ph]].
%
\bibitem{ALGEBRA}
  J.~Bl\"umlein,
  Comput.\ Phys.\ Commun.\  {\bf 159} (2004) 19
  [arXiv:hep-ph/0311046].
%
\bibitem{BRK}
  J.~Bl\"umlein, V.~Ravindran,
  Nucl.\ Phys.\  {\bf B716 } (2005)  128
  [hep-ph/0501178];
  Nucl.\ Phys.\  {\bf B749 } (2006)  1
  [hep-ph/0604019];\\
  J.~Bl\"umlein, S.~Klein,
  [arXiv:0706.2426 [hep-ph]].
%
\bibitem{TFORM}
  M.~Tentyukov and J.~A.~M.~Vermaseren,
  Comput.\ Phys.\ Commun.\  {\bf 181} (2010) 1419
  [hep-ph/0702279];\\
  J.~A.~M.~Vermaseren,
  Nucl.\ Phys.\ Proc.\ Suppl.\  {\bf 205-206} (2010) 104
  [arXiv:1006.4512 [hep-ph]].
%
\bibitem{SIGMA}
C. Schneider, J. Symbolic Comput. 43 (2008) 611, \newblock [arXiv:0808.2543v1]; Ann. Comb. 9 (2005) 75;
J. Differ. Equations Appl. 11 (2005) 799; Ann. Comb.  {\bf 14} (4) (2010),
[arXiv:0808.2596]; Proceedings of the Workshop ``Motives, Quantum Field Theory, and Pseudodifferential 
Operators'', held at the Clay Mathematics
       Institute, Boston University, June 2--13, 2008, 
       Clay Mathematics Proceedings {\bf 12} (2010) 285-308, Eds. A.~Carey,
       D.~Ellwood, S.~Paycha, S. Rosenberg;
S\'em.~Lothar. Combin. 56 (2007) 1, Article B56b,  Habilitationsschrift JKU Linz (2007)
and references therein;\\
  J.~Ablinger, J.~Bl\"umlein, S.~Klein, C.~Schneider,
  Nucl.\ Phys.\ Proc.\ Suppl.\  {\bf 205-206 } (2010)  110
  [arXiv:1006.4797 [math-ph]].
%
\bibitem{QEXP}
  R.~Harlander, T.~Seidensticker, M.~Steinhauser,
  Phys.\ Lett.\  {\bf B426 } (1998)  125, 
  [hep-ph/9712228];\\
  T.~Seidensticker,
  [hep-ph/9905298].
%
\bibitem{BVN}
  J.~Bl\"umlein, W.~L.~van Neerven,
  Phys.\ Lett.\  {\bf B450 } (1999)  417
  [hep-ph/9811351].
%
\bibitem{DO}
  M.~Gl\"uck, E.~Reya, M.~Stratmann,
  Nucl.\ Phys.\  {\bf B422 } (1994)  37. 
%
\bibitem{ABKM}
  S.~Alekhin, J.~Bl\"umlein, S.~Klein, S.~Moch,
  Phys.\ Rev.\  {\bf D81 } (2010)  014032.
  [arXiv:0908.2766 [hep-ph]].
%
\bibitem{ABHKSW}
J. Ablinger, J. Bl\"umlein, A. Hasselhuhn, S. Klein, C. Schneider, and F. Wi\ss{}brock, in
preparation.
%
\bibitem{Buza:1996wv}
  M.~Buza, Y.~Matiounine, J.~Smith and W.~L.~van Neerven,
  Eur.\ Phys.\ J.\ C {\bf 1} (1998) 301
  [hep-ph/9612398].
%
\bibitem{BHS}
J. Bl\"umlein, A. Hasselhuhn, and C. Schneider, in preparation.
%
\bibitem{Remiddi:1999ew}
  E.~Remiddi and J.~A.~M.~Vermaseren,
  Int.\ J.\ Mod.\ Phys.\ A {\bf 15} (2000) 725
  [hep-ph/9905237].
%
\bibitem{Ablinger:2011te}
  J.~Ablinger, J.~Bl\"umlein and C.~Schneider,
  J. Math. Phys. {\bf 52} (2011) 102301  [arXiv:1105.6063 [math-ph]].
%
\bibitem{HSUM}
  J.~A.~M.~Vermaseren,
  Int.\ J.\ Mod.\ Phys.\ A {\bf 14} (1999) 2037
  [hep-ph/9806280];\\
  J.~Bl\"umlein and S.~Kurth,
  Phys.\ Rev.\ D {\bf 60} (1999) 014018
  [hep-ph/9810241].
%
\bibitem{MZV}
  J.~Bl\"umlein, D.~J.~Broadhurst and J.~A.~M.~Vermaseren,
  Comput.\ Phys.\ Commun.\  {\bf 181} (2010) 582
  [arXiv:0907.2557 [math-ph]].
%
\bibitem{ADISS}
J. Ablinger, PhD Thesis, JKU Linz, April 2012. 
\end{thebibliography}
\end{document}